\documentstyle[12pt,epsf]{article}

\advance \topmargin by -\headheight
\advance \topmargin by -\headsep
     
\evensidemargin \oddsidemargin
\begin{document}

\title{\Large\bf Inverse meson mass ordering in color-flavor-locking
phase of high density QCD: erratum}

\author{D.T.~Son$^{1,3}$ and M.A.~Stephanov$^{2,3}$\\
{\small\em $^1$ Physics Department, Columbia University, New York, NY 10027} \\
{\small\em $^2$ Department of Physics, University of Illinois,
Chicago, IL 60607-7059}\\
{\small\em $^3$ RIKEN-BNL Research Center, Brookhaven National Laboratory,
Upton, NY 11973}}
\date{April 2000}
\maketitle

\begin{abstract}
We correct a mistake in
the calculation of meson masses at large baryon chemical
potential $\mu$ made in Ref.\ \cite{paper}. 
\end{abstract}

The values of the meson masses reported in Ref.\ \cite{paper}
are not correct. 
The correct answer for
the meson masses and mixings
are given by Eqs.\ (45) and (46) in Ref.\ \cite{paper} 
with $C=2c/f_\pi^2$, and
\begin{equation}
  c = {3\Delta^2\over2\pi^2},
\label{c}
\end{equation}
where $\Delta$ is the value of the superconducting gap at the Fermi
surface.  To leading order in perturbation theory, the constant
denoted in Ref.\ \cite{paper} as $c'$ vanishes.
Parametrically, the meson masses are of order $m_q\Delta/\mu$ instead
of $m_q$ as reported in Ref.\ \cite{paper}.  Moreover, since $c'$
vanishes to the leading order, a truly inverse mass hierarchy emerges
instead of a partial one previously reported.
This means the inverse hierarchy
anticipated by the 
heuristic argument following Eq.\ (46) in Ref.\ \cite{paper} 
is indeed realized.

Technically, the mistake of our previous calculation of the masses
came from the failure to take into account the fact that the quark
mass insertion in a quark line changes the quark mass shell from
$E=|\mbox{\boldmath$p$}|-\mu$ to $E=-|\mbox{\boldmath$p$}|-\mu$
~\cite{Hong}.  As a
result, an almost on-shell quark near the Fermi surface becomes an
off-shell particle with virtuality $2\mu$.

\begin{figure} \centering
$$
      \def\epsfsize #1#2{0.5#1}
      \epsfbox{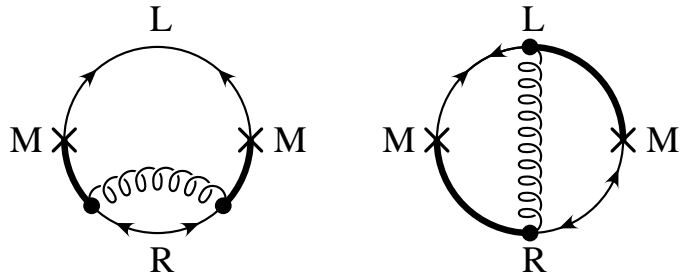}
$$
\caption[]{
These two diagrams (together with their parity and
charge conjugation mirrors) contribute to the shift of vacuum energy
to the leading order in $g$. Thin lines denote quarks which are
almost on-shell near the Fermi surface (virtualities of order
$\Delta$). Thick lines correspond to the off-shell quarks (virtualities
close to $2\mu$).}
\label{fig:diagrams}
\end{figure}

The correct computation of the constants $c$ and $c'$ follows the same
methodology of Ref.\ \cite{paper} of matching vacuum energies and
involves the evaluation of two Feynman diagrams in
Fig.\ \ref{fig:diagrams}, which both make leading-order contributions
to the shift of the vacuum energy.
Note that the gap insertions, which reverse the direction of the quark
lines in Fig.\ \ref{fig:diagrams}, occur only on the propagators which
are near the particle mass shell.  The lower part of the first diagram
between $M$ insertions corresponds to the one-loop quark 
self-energy and is equal, by the gap
equation, to a quantity commonly referred to as the ``anti-particle
gap''.  The latter is, however, not gauge invariant even to
leading order in perturbation theory.  Consequently, the first diagram
is not gauge-fixing independent on its own.  However, the sum of the
two diagrams is indeed gauge-fixing independent, which is a
consequence of Ward-Takahashi identity and can be verified by a direct
calculation.

The following comments concerning the result (\ref{c}) are in order.
The contribution of the {\em magnetic} gluon exchange cancels
between the two diagrams. This happens because emission of a
magnetic gluon by an on-shell quark is suppressed by parity
conservation.  As a result, $c$ is proportional
to $\Delta^2$ instead of $\Delta^2\log(\mu/\Delta)$. The remaining electric
gluon exchange is suppressed at small gluon momenta because
of the angular momentum conservation and the requirement
that the helicity flip must occur in the diagram.  This removes
a collinear divergence which would have given rise to an extra factor of
$1/\log g$ \cite{paper2}.

\end{document}